\documentclass{JHEP}
\usepackage{epsfig}
\usepackage{amssymb}

\newcommand{\sect}[1]{\setcounter{equation}{0}\section{#1}}
\newcommand{\subsect}[1]{\subsection{#1}}

\newcommand{\be}{\begin{equation}}
\newcommand{\ee}{\end{equation}}
\newcommand{\bea}{\begin{eqnarray}}
\newcommand{\eea}{\end{eqnarray}}
\newcommand{\p}{\partial}

\newcommand{\msc}[1]{\mbox{\scriptsize #1}}
\newcommand{\cQ}{{\cal Q}}
\newcommand{\NS}{\mbox{NS}}

\newcommand{\R}{\mbox{R}}

\newcommand{\sNS}{\msc{NS}}

\newcommand{\cN}{{\cal N}}

\newcommand{\dr}{{\rm d}}
\newcommand{\pp}{{\rm p}}
\newcommand{\dket}[1]{{\left.\left|#1\right\rangle\right\rangle}}

\newcommand{\lb}{\lbrack}
\newcommand{\rb}{\rbrack}
\newcommand{\dsp}{\displaystyle}
\newcommand{\ch}[2]{\mbox{ch}^{#1}_{#2}}

\renewcommand{\th}{{\theta}}

\newcommand{\f}{\frac}
\preprint{ROM2F/2004/33\\ SNUST 041103\\ UT-04-31 \\
{\tt hep-th/0412043}}
\title{Rolling Down the Throat in NS5-brane Background:\\
The Case of Electrified D-Brane}
\author{Yu Nakayama${}^{a}$, Kamal L. Panigrahi ${}^b$, Soo-Jong Rey ${}^{c}$ \&
Hiromitsu Takayanagi ${}^{a}$\\
~~~~~~~~~~~\\
${}^a$ Department of Physics, Faculty of Science, University of Tokyo\\
Hongo 7-3-1, Bunkyo-ku, Tokyo 113-0033  {\rm JAPAN}\\
${}^b$ Departimento de Fisica, Universita' di Roma  \&
INFN, Sezione di Roma ``Tor Vergata" \\
Via della Ricerca Scientifica 1, Roma 00133 {\rm ITALY}\\
${}^c$ School of Physics \& BK21 Physics Division\\
Seoul National University, Seoul 151-747 {\rm KOREA}\\
~~~~~~~~~~~~~~~~\\
\email{\tt nakayama@hep-th.phys.s.u-tokyo.ac.jp, \hskip0.5cm
Kamal.Panigrahi@roma2.infn.it \hskip0.5cm sjrey@phya.snu.ac.kr,
\hskip0.5cm hiro@hep-th.phys.s.u-tokyo.ac.jp} }
\abstract{
We study rolling radion dynamics of electrified D-brane in NS5-brane
background, both in effective field theory and in full open string
theory. We construct exact boundary states and, from them, extract
conserved Noether currents. We argue that T-duality and Lorentz
boost offer an intuitive approach. In the limit of large number of
NS5-branes, both boundary wave functions and conserved currents are
sharply peaked and agree with those deduced from the effective field
theory. As the number of NS5-branes is reduced, width around the
peak becomes wider by string corrections. We also study radiative
decay process. By applying Lorentz covariance, we show how the decay
of electrified D-brane is related to that of bare D-brane. We
compute spectral moments of final state energy and winding quantum
number. Using Lorentz covariance argument, we explain in elementary
way why winding quantum number should be included and derive rules
how to do so. We conclude that Kutasov's ``geometric realization"
between radion rolling dynamics and tachyon rolling dynamics holds
universally, both for bare and electrified D-branes.}
\begin{document}
\section{Introduction}
An outstanding open problem in string theory is to formulate a
framework for addressing string dynamics in time-dependent and
cosmological backgrounds. By establishing such a framework, one
hopes to understand better not only string theory itself but also
conceptual issues in quantum gravity and quantum cosmology. Prompted
by ground-breaking work of Sen \cite{sen-RT}, there has been
considerable progress in understanding time-dependent phenomena in
the open string theory: decay of unstable D-brane or brane-antibrane
pair is describable by rolling and condensation of open string
tachyon. Sen further argued \cite{sen-TM} that, as the tachyon rolls
down to potential minimum, an unstable D-brane or a brane-antibrane
pair decays to a system consisting of pressureless dust called
``tachyon matter" yet devoid of any obvious open string excitation.

Formation of non-threshold bound-state among branes is another
situation that time-dependent string dynamics is involved. A
well-known case is formation of a supersymmetric bound-state $(p,q)$
string \cite{witten, narain} when a probe BPS D-string of charge
$(p,0)$ falls into a target BPS F-string of charge $(0,q)$
\cite{leerey, bakreyyee}. Another case is formation of a
supersymmetric bound-state of D-brane and NS5-brane \cite{seiberg}
when a probe BPS D-brane falls into a target BPS NS5-brane
\cite{Kutasov:2004dj}. The supergravity background of NS5-brane
\cite{Rey:1989xj, Callan:1991dj} becomes in the near horizon limit
an exactly solvable conformal field theory (CFT) \cite{ReyCFT,
ReyCFT2, CHS}, known as the ``throat geometry".  In both cases,
radion rolling dynamics is describable in terms of the worldvolume
theory in the background of target F-string or NS5-brane.
Interestingly, in the worldvolume description, radion rolling
dynamics of BPS D-branes resembles tachyon rolling dynamics of
unstable D-branes or brane-antibrane pair. In particular, for
appropriate background and regime of target NS5-branes, the works
\cite{Kutasov:2004dj, Kutasov2} noted that the radion effective
action takes exactly the same functional form as the tachyon
effective action for unstable D-branes and, from this observation, proposed to view the
radion rolling dynamics as a sort of ``geometric realization" of
tachyon rolling dynamics for unstable D-branes.

Is such ``geometric realization" an artifact of low-energy effective
approach or does it extend to full-fledged string theoretic analysis?
This question was addressed in \cite{Nakayama:2004yx}. There, by
utilizing the fact that the ``throat geometry" is an exactly
solvable CFT, boundary state of the rolling radion was constructed as
a counterpart of boundary state for the rolling tachyon
\cite{Sen:2004nf}. From them, positive conclusion was drawn that the
``geometric realization" extends to the full string theory beyond
the supergravity limit. See also \cite{Sahakyan:2004cq}.\footnote{Related works can be found in
\cite{Yavartanoo:2004wb,Panigrahi:2004qr,Ghodsi:2004wn,Nakayama:2004at,
Saremi:2004yd,Kluson:2004xc,Toumbas:2004fe,Kluson:2004yk,Thomas:2004cd}.}
More specifically, boundary state of the infalling D-brane was
constructed by relating the CFT of the ``throat geometry"
\cite{ReyCFT,  ReyCFT2, CHS} to $\mathcal{N}=2$ Liouville theory\footnote{See \cite{Nakayama:2004vk} for review on the relation.}
and then by performing a suitable Wick rotation to so-called
noncompact class-2 brane  in the $\mathcal{N} = 2$ Liouville theory
\cite{Nakayama:2004yx}. The rolling radion boundary state
constructed in \cite{Nakayama:2004yx} facilitated to study the
dynamics exactly and to compare it with the dynamics of rolling
tachyon for unstable D-brane. In particular, the investigation of
decay rate to closed strings indicated that both rolling radion and
rolling tachyon share exactly the same properties. Such universality
is newly emergent feature that may lead to breakthroughs for better
understanding of time-dependent phenomena in open string theory.

The purpose of this work is mainly twofold. First, this work aims at
investigating aspects of the rolling radion boundary state
constructed in \cite{Nakayama:2004yx} and to put it into a further
consistency check. We shall do so by turning on homogeneous electric
field on the worldvolume of the rolling Dp-brane in the direction
parallel to the background NS5-brane worldvolume. This is the
rolling radion counterpart to the tachyon rolling of unstable
D-brane with worldvolume electric field \cite{Mukhopadhyay:2002en,
Rey:2003zj}. As emphasized in \cite{Rey:2003zj}, electrifying a
Dp-brane is equivalent in an appropriate limit to T-dual of
boosting the D(p-1)-brane along the electric field direction. It
is then intuitively clear that physical observables ought to
transform covariantly under such Lorentz boost and T-duality map
\cite{toappear}. Adopting the same strategy as \cite{Rey:2003zj}, we
shall construct boundary state of rolling radion with electric
field, study decay rate to closed strings, and compare the result
with that expected from intuitive approach based on Lorentz boost
and T-duality map. Agreement between the two methods would serve as a
consistency check that the boundary state constructed in
\cite{Nakayama:2004yx} is indeed a correct one. Second, this work
aims at studying whether and, if so, how F-strings of macroscopic
size are formed as probe Dp-brane falls into the target NS5-brane.
In an appropriate limit, the aforementioned Lorentz invariance
argument asserts that such F-strings are certainly formed. An
interesting question would then be details of spectral distribution
of the formed F-strings. In fact, this issue is precisely the
counterpart of F-string formation during tachyon rolling of unstable
D-brane \cite{Rey:2003zj,Sen-99,GHY}, an issue which was revisited
in \cite{Sen:2003xs} in a different context.

Investigation of tachyon rolling for electrified D-brane was made
precise in \cite{Mukhopadhyay:2002en,Rey:2003zj} by constructing
relevant boundary state and then comparing the dynamics
against that deduced from effective field theory approach. As
emphasized in \cite{Rey:2003zj}, final decay product consists of
fundamental strings in addition to ``tachyon matter", a fact which
again can be readily understood from the Lorentz invariance. We
shall follow the same strategy and construct the boundary state of
rolling radion for an electrified D-brane. We first study the
Born-Infeld effective action of a rolling D-brane in the presence of electric
field and obtain D-brane trajectory and energy-momentum tensor. Then we extend the
analysis done in \cite{Nakayama:2004yx} and construct the boundary
states for the rolling D-brane in the presence of the constant
electric field. Our construction of boundary states is based on the
prescription proposed in \cite{Rey:2003zj} in the context of
boundary states of the rolling tachyon in the presence of constant electric
field. We will see that this prescription reproduces
correct energy-momentum tensor and D-brane trajectory in the
supergravity limit, $N \rightarrow \infty$.   We also discuss
quantum aspects by studying the closed string emission rate from the
rolling D-brane. By applying the Lorentz boost and using Lorentz covariance
of spectral observables, we prove in elementary way that winding quantum numbers
are to be included. We find that universal feature of the decay
process is recovered once we include winding modes.

This paper is organized as follows. In section \ref{2}, as
a prelude, we study the effective field theory approach to radion
rolling dynamics of electrified D-brane in the NS5-brane background.
We obtain classical trajectory and conserved current tensors of the
electrified D-brane. In section \ref{3}, we propose the radion
rolling boundary states of an electrified D-brane by utilizing the
prescription introduced in \cite{Rey:2003zj}. In section \ref{4}, we
extract conserved currents directly from the boundary states
proposed in section \ref{3}. We first review how to read the
conserved currents for radion rolling of a bare D-brane. Then we
show that conserved currents obtained directly from the boundary
states constructed in section \ref{3} agree in the supergravity
limit with those obtained from the effective field theory approach.
In section \ref{5}, we study radiation of the bing energy into
closed string as the electrified D-brane rolls down to the
NS5-brane. We then compare it with the situation for tachyon rolling
of unstable D-brane. By appealing to underlying Lorentz invariance,
we show that electrifying rolling D-brane is simply accounted for by
taking into account of closed strings with nonzero winding number.
Section \ref{6} is devoted to conclusion and discussions for points
worthy of further investigation.

\textbf{Note Added}: After the completion of this work, we found the
work \cite{Chen:2004vw} posted in arXiv.org, which partially
overlaps with our work, in which the electrified D-brane
boundary states for the rolling radion is obtained by
Lorentz boosting zero-mode boundary wave function. We shall
show in this work that proper treatment of the oscillator transformation
based on the $\cN=2$ Liouville boundary states is imperative
for obtaining correct conserved currents from the boundary states.

\sect{Prelude: Effective Field Theory Approach}\label{2}

We shall begin with effective field theory approach to radion
rolling of electrified D-brane in the background of NS5-branes.
The supergravity background of a stack of $N$ coincident NS5-branes
is given by \cite{Rey:1989xj,CHS,Horowitzstrominger}: \bea ds^2_{\rm
string} &=& \eta_{\alpha \beta} ~dx^{\alpha}dx^{\beta}
+ H (x^n)~\delta_{mn}~ dx^m dx^n\ ,\\
e^{2\Phi} &=& g^2_{\rm st} H(x^n),  \qquad \quad H_{mnp} =
-{\epsilon^q}_{mnp}\p_q \Phi \ ,  \nonumber \eea where $\alpha, \beta
= 0,...,5;\>\> m,n= 6,...,9$ refer to directions parallel and
transverse, respectively, to the NS5-brane worldvolume and
\bea H (x^n)= 1 + {{N\alpha'}\over r^2} \nonumber  \eea
is the harmonic function in the direction transverse to the
NS5-branes. Consider a Dp-brane extended parallel to the
NS5-branes. D$p$-brane dynamics in the transverse space was studied
in terms of Born-Infeld action in \cite{Kutasov:2004dj}. There, it
was shown that absorption of D-brane to NS5-branes resembles decay
of unstable D-brane in flat space via tachyon condensation. In the
NS5-brane background, the string coupling becomes large and
semiclassical analysis would typically break down. An interesting
point, as observed in \cite{Kutasov:2004dj}, is that there exists a
range of D-brane energy for which a significant part of the
dynamical process occurs when the string coupling is rather weak and
the perturbative string theory approach would still be applicable.

\subsection{Effective action and classical trajectory}
We shall now extend the effective field theory analysis of
\cite{Kutasov:2004dj} to electrified D-brane --- a D-brane with
electric flux turned on. So consider $B_{01} = - B_{10} =
\varepsilon$, with all other components of $B_{\mu\nu}$ being zero.
The Born-Infeld action of Dp-brane now reads\footnote{We denote
Dp-brane worldvolume parameters as $\sigma^\mu$
($\mu=0,1,\cdots, \pp$) and tension as $\tau_\pp \equiv g_{\rm st}^{-1}
(2 \pi)^{-\pp} \alpha'^{-(\pp+1)/2}$. } :
\bea S_{\rm p} = - \tau_{\pp} g_{\rm st} \int \dr^{\pp+1}
\sigma \,  e^{-\Phi}\sqrt{-\det(X^*[G+ B]_{\mu\nu})}\ ,
\label{orDBI} \eea 
where $X^*[G+ B]$ denotes pullback to D$p$-brane worldvolume: 
\bea X^*[G]_{\mu\nu} &=& {{\partial
X^{A}}\over{\partial \xi^{\mu}}} {{\partial X^{B}}\over{\partial
\xi^{\nu}}}g_{AB} \qquad \mbox{and} \qquad X^*[B]_{\mu\nu} =
{{\partial X^{A}}\over{\partial \xi^{\mu}}} {{\partial
X^{B}}\over{\partial \xi^{\nu}}}b_{AB} \qquad (A, B = 0, \cdots, 9)
\, . \nonumber \eea
We shall fix worldvolume reparametrization invariance by static
gauge: $\sigma^{\mu} = x^{\mu}$. The pullback is then given by
\bea A_{\mu \nu} \equiv X^*[G+ B]_{\mu \nu} = g_{\mu \nu} + b_{\mu
\nu} + H (R) \p_{\mu} R \p_{\nu} R \ . \nonumber \eea
For rigid motion of a Dp-brane, denoting its transverse position
as $R(t)$ and spatial volume as $V_{\rm p}$, the Born-Infeld action
Eq.(\ref{orDBI}) is reduced to
\bea
S_{\rm p} &=& - \tau_{\rm p} V_{\rm p} \int{\dr t \, \sqrt{{{1-\varepsilon^2}\over H(R)}
- \Big({\dr R \over \dr t}\Big)^2}} \nonumber \\
&=& - \tau_\pp V_\pp \int \dr s \sqrt{ {1 \over H(R)} - \Big({\dr R
\over \dr s}\Big)^2 }\, . \label{dbi-t} \eea
In the second line, we recasted the action in a suggestive form by
introducing ``boosted time" $s \equiv \sqrt{1 -\varepsilon^2} t$. We
are interested in Dp-brane dynamics primarily in the the ``throat
region", viz. $ R \ll \sqrt{N \alpha'}$. In this region,  the action
Eq.(\ref{dbi-t}) is further simplified to\footnote{In the following
expressions, we have rescaled $\tau_\pp \rightarrow \tau_\pp g_{\rm st}$
and $R\rightarrow g^{-1}_{\rm st} R$.}
\bea S_\pp = - \tau_\pp V_\pp \int{\dr t \, {R\over{\sqrt {N \alpha'}}}
\sqrt{1-\varepsilon^2 - N \alpha' \left({\dr \over \dr t} ~{\log {R
\over \sqrt{N \alpha'}} } \right)^2}} \ . \label{dbi-nh} \eea
Change the dynamical variable as
\bea \exp \Big({\phi \over{\sqrt {N\alpha'}}}\Big) \equiv
{R\over{\sqrt{N\alpha'}}} \ll 1 \, , \nonumber \eea
and the action Eq.(\ref{dbi-nh}) now reads
\bea S_\pp = -\tau_\pp V_\pp \int{\dr t \, V(\phi) \sqrt{1-\varepsilon^2 -
\dot{\phi}^2}},  \qquad \mbox{where} \qquad V(\phi) = \exp
\Big({\phi \over{\sqrt {N\alpha'}}} \Big)\ . \label{dbi-new} \eea
One readily finds the classical trajectory as
\bea \phi(t) = - \sqrt{N\alpha'}~\ln \left( {\tau_\pp V_\pp \over E}
\cosh {\sqrt{1-\varepsilon^2}t\over{\sqrt {N\alpha'}}} \right) \ ,
\label{traj} \eea
where $E$ is the conserved total energy measured in unit of the
``boosted time" $s = \sqrt{1 - \varepsilon^2} \, t$. Evidently, both
the Born-Infeld action Eq.(\ref{dbi-t}) and the classical trajectory
Eq.(\ref{traj}) are simply those of Dp-brane in NS5-brane
background but in terms of the ``boosted time" $s= \sqrt{1 -
\varepsilon^2} \, t$. This statement will become clearer in the next
section where we construct exact radion rolling boundary state of
the Dp-brane via a chain of mapping involving Lorentz boost and
T-dualities.

It is also interesting to compare the situation with the trajectory
with nonzero angular momentum $L$ \cite{Kutasov:2004dj}: \bea
\phi(t) = - \sqrt{N\alpha'}~\ln
\left(\frac{\tau_\pp V_\pp}{E\sqrt{1-\frac{L^2}{N\alpha'
E^2}}}\cosh\frac{t}{\sqrt{N\alpha'}}\sqrt{1-\frac{L^2}{N \alpha'
E^2}}\right) \ . \nonumber \eea
Analogy with Eq.(\ref{traj}) is evident: T-dual of ``internal"
Lorentz-boost is replaced by ``rotational" Lorentz-boost.
\subsect{Conserved current tensors}

Conserved Noether currents provide useful probe for analyzing
D-brane dynamics. For later comparison, we shall now derive
energy-momentum and string current tensors for the electrified
D-brane rolling along the classical trajectory Eq.(\ref{traj}).
Varying the Born-Infeld action Eq.(\ref{orDBI}) with respect to
background closed string fields, we obtain\footnote{From now on, we
set $\alpha' = 2$ for our convenience.}
\bea T_{\mu \nu} &\equiv& 2 {\delta S_\pp \over \delta
g^{\mu \nu}} = -\frac{\tau_\pp g_{\rm st} }{2}e^{-\Phi} \sqrt{-{\rm
det} A} (A^{-1})_{(\mu \nu)} \nonumber \\
Q_{\mu \nu} &\equiv& 2 {\delta S_\pp \over \delta b^{\mu \nu}}=
-\frac{\tau_\pp g_{\rm st} }{2}e^{-\Phi} \sqrt{-{\rm det} A}
(A^{-1})_{[\mu \nu]} \ . \nonumber
\eea
Explicitly,
\bea
T_{00} &=& +\frac{\tau_\pp}{\sqrt{H}} \frac{1}{\sqrt{1- \varepsilon^2
-H\dot{X}^m \dot{X}^m}}\ , \nonumber \\
T_{11} &=& -\frac{\tau_\pp}{\sqrt{H}} \frac{1-H\dot{X}^m
\dot{X}^m}{\sqrt{1-\varepsilon^2
- H\dot{X}^m \dot{X}^m}}\ , \nonumber \\
Q_{01} &=& +\frac{\tau_\pp}{\sqrt{H}}
\frac{\varepsilon}{\sqrt{1-\varepsilon^2
- H\dot{X}^m \dot{X}^m}}\ , \nonumber \\
T_{ij} &=& -\frac{\tau_\pp}{\sqrt{H}} \sqrt{1-\varepsilon^2-H\dot{X}^m
\dot{X}^m}~\delta_{ij} \ , \qquad\qquad (i,j = 2,...,\pp) \, \nonumber
\eea
and all other components vanish. In these expressions, we also
suppressed delta function factors localizing the D-brane on the
classical trajectory Eq.(\ref{traj}). They are all conserved Noether
currents.

In the near-horizon limit, from the Born-Infeld action
Eq.(\ref{dbi-new}), the conserved current tensors are reduced to
\bea T_{00} &=& +\tau_\pp e^{\frac{\phi}{\sqrt{2N}}}
\frac{1}{\sqrt{1-\varepsilon^2 - \dot{\phi}^2}} \ ,
\nonumber \\
T_{11} &=& -\tau_\pp e^{\frac{\phi}{\sqrt{2N}}} \frac{1-\dot{\phi}^2}
{\sqrt{1-\varepsilon^2-\dot{\phi}^2}} \ , \nonumber \\
Q_{01} &=& +\tau_\pp e^{\frac{\phi}{\sqrt{2N}}} \frac{\varepsilon}
{\sqrt{1-\varepsilon^2-\dot{\phi}^2}} \ , \nonumber \\
T_{ij} &=& -\tau_\pp e^{\frac{\phi}{\sqrt{2N}}}
\sqrt{1-\varepsilon^2-\dot{\phi}^2}~\delta_{ij} \qquad \qquad  (i,j
= 2,...,\pp)\ . \label{emts} \eea
Reinstating delta function factors for $\phi(t)$ and thus imposing
the classical trajectory Eq.(\ref{traj}), we obtained the following
expressions for the conserved current tensors\footnote{Other
components are deducible from those in Eq.(\ref{emts}) via modified current
conservation to be discussed later. See Eq.(\ref{conse}).}:
\bea
T_{00} &=& +\frac{E}{V_\pp} \gamma \, ~\delta(\phi-\phi(t)) \ , \nonumber \\
T_{11} &=& -\frac{E}{V_\pp} \gamma \, \left( \mbox{sech}^2
\Big(\frac{\gamma^{-1}t}{\sqrt{2N}}\Big) +\varepsilon^2\tanh^2
\Big(\frac{\gamma^{-1}t}{\sqrt{2N}}\Big)\right)\delta(\phi-\phi(t))\ , \nonumber \\
Q_{01} &= & +\frac{E}{V_\pp}~\varepsilon~\gamma~\delta(\phi-\phi(t))\ , \nonumber \\
T_{ij} &=& -\delta_{ij} \, \frac{E}{V_\pp}~\gamma^{-1} \mbox{sech}^2
\Big(\frac{\gamma^{-1}t}{\sqrt{2N}}\Big) \delta(\phi-\phi(t))\ , \nonumber \\
T_{0\phi}&=& +\frac{E}{V_\pp}~{\tanh}\Big(\frac{\gamma^{-1}t}{\sqrt{2N}}\Big)
~\delta(\phi-\phi(t)) \nonumber\ , \\
T_{\phi \phi} &=& +\frac{E}{V_\pp}~\gamma^{-1}~{\tanh^2}
\Big(\frac{\gamma^{-1}t}{\sqrt{2N}}\Big)~\delta(\phi-\phi(t))\ , \nonumber \\
Q_{\phi 1}
&=&+\frac{E}{V_\pp}~\varepsilon~{\tanh}\Big(\frac{\gamma^{-1}t}{\sqrt{2N}}\Big)
~\delta(\phi-\phi(t))\ . \label{emtens} \eea
Here, $\gamma \equiv 1/\sqrt{1 - \varepsilon^2}$. At asymptotic
future infinity, $t \rightarrow \infty$, all current components
vanish except the first three, viz. $T_{00}, T_{11}, Q_{01}$. We
readily notice that these currents consist of the two parts ---
pressureless dust referred as ``radion matter" (rolling radion
counterpart to the tachyon matter) and pressure-carrying F-string
dust. Indeed, in the limit $\varepsilon \to 0$, the classical
trajectory Eq.(\ref{traj}) and the conserved currents
Eq.(\ref{emtens}) are reduced respectively to\footnote{This agrees
with the result obtained in \cite{Sahakyan:2004cq}.}
\bea \phi_0(t)=-\sqrt{2N}\ln \Big(\f{\tau_\pp V_\pp}{E}\cosh
(\frac{t}{\sqrt{2N}}) \Big)
\nonumber \eea
and
\bea
T_{00}&=& +\f{E}{V_\pp}\delta(\phi-\phi_0(t))\ , \nonumber \\
T_{0\phi}&=& +\f{E}{V_\pp} \tanh\Big(\frac{t}{\sqrt{2N}}\Big)
\delta(\phi-\phi_0(t)) \ ,
\nonumber \\
\quad T_{ij}&=&-\f{E}{V_\pp} \mbox{sech}^2
\Big(\frac{t}{\sqrt{2N}}\Big)\delta(\phi-\phi_0(t))\delta_{ij}
\qquad\qquad  (i,j = 1,...,\pp) \ .\nonumber \eea
In this case, the pressure vanishes monotonically as $t \rightarrow
\infty$, yielding a pressureless ``radion matter". Then, the rest of
the contribution arising at nonzero $\varepsilon$ ought to be
attributed to electifying the Dp-brane. Such two-component fluid
behavior would also have been anticipated since turning on electric
field is T-dual in a suitable (de)compactification limit to Lorentz
boost of D(p-1)-brane in a direction parallel to NS5-brane with
constant velocity $\varepsilon$.

Conserved currents analyzed above indeed behave the same as those
for unstable D-brane \cite{Mukhopadhyay:2002en, Rey:2003zj}, thus
supporting expectation that the ``geometric realization" would
persist to hold to the situation where the D-brane is electified. In
the next section, we shall show that the ``geometric realization"
can also be seen from full-fledged string theory construction in
terms of relevant boundary states.
\sect{Boundary States for Electrified D-Brane}\label{3}
String theoretic account for radion rolling dynamics of electrified
D-brane requires construction of an appropriate boundary state. We
shall do so in this section. We first recapitulate aspects of the
rolling radion boundary state of a bare D-brane constructed in
\cite{Nakayama:2004yx} relevant for foregoing discussions. The
construction proceeded by performing an appropriate Wick rotation of
the hairpin D-brane \cite{Ribault:2003ss,Lukyanov:2003nj}, which
corresponds to so-called class-2 brane in the $\mathcal{N} = 2$
Liouville theory \cite{Eguchi:2003ik}.
\subsect{Bare D-brane rolling in ``throat geometry"} In the near
horizon limit, stack of $N$ NS5-branes develops so-called ``throat
geometry". String theory in this background is described by the
superconformal field theory \cite{ReyCFT,ReyCFT2,CHS}
\begin{eqnarray}
\mathbb{R}^{5,1} \times \mathbb{R}_{\phi} \times SU(2)_{N-2} \cong
\mathbb{R}^{5,1} \times \frac{\left\lb \mathbb{R}_{\phi} \times
\mathbb{S}^1_Y \right\rb \times
  M_{N-2}}{\mathbb{Z}_{N}}~,
\label{CHS}
\end{eqnarray}
where $M_{k}$ denotes the $\cN=2$ minimal model with level $k$ and
central charge $\hat{c}=k/(k+2)$, and $\mathbb{R}_{\phi} \times
\mathbb{S}^1_Y$ denotes the $\cN=2$ Liouville theory with
$\hat{c}=1+\cQ^2 = 1+\frac{2}{N}$. The $\mathbb{Z}_N$-orbifolding
acts as the Gliozzi-Scherk-Olive (GSO) projection and enforces
integrality of the total $\cN=2$ $U(1)$-charge.

To construct the boundary state for noncompact D-brane and to enable
appropriate Wick rotation into the boundary state of the sought-for
radion-rolling D-brane, we consider another $\cN =2$ Liouville
system as part of the fundamental building block. For this, we take
the $U(1)$ direction of the $\cN = 2$ Liouville theory out of the
noncompact $ X \in \mathbb{R}^{5,1}$ (rather than out of the $SU(2)$
part, as usually done). This result in $\cN=2$ superconformal
algebra generated by the currents
\bea
 T ~&=&-\frac{1}{2}(\partial X)^2
-\frac{1}{2}(\partial \phi)^2 -\frac{\cQ}{2}\partial^2\phi
-\frac{1}{2}(\Psi^+\partial \Psi^- -\partial \Psi^+ \Psi^-) \nonumber \\
  G^{\pm} &=& -\Psi^{\pm}(i\partial X \pm \partial \phi )
\mp {\cQ}\partial \Psi^{\pm} \nonumber \\
J ~&=& \Psi^+\Psi^- - \cQ i\partial X~~, \label{SCA-L} \eea
where $\dsp X(z)X(0)\sim -\ln z $, $\dsp \phi(z)\phi(0)\sim -\ln z$,
$\dsp \Psi^{\pm}(z)\Psi^{\mp}(0)\sim \frac{1}{z}$, $\dsp
\Psi^{\pm}(z)\Psi^{\pm}(0) \sim 0 $, and $\dsp
\Psi^{\pm}=-\frac{1}{\sqrt{2}}(\psi^X\pm i\psi^{\phi})$ are free
fields. The $\cN=1$ supercurrent is embedded as $G =(G^+ +
G^{-})/\sqrt{2}$. The linear dilaton background runs into strong
string coupling singularity. We resolve the singularity by turning
on an appropriate $\cN=2$ Liouville potential. There are in fact
several viable choices of the potential, all compatible with the
${\cal N}=2$ superconformal currents Eq.(\ref{SCA-L}). For the
present system, adopting argument of \cite{Nakayama:2004yx},
nonchiral Liouville potential is better suited. However, for
foregoing considerations, precise form of the Liouville potential is
largely irrelevant, so we will not specify the choice explicitly.
Furthermore, since dependence on the cosmological constant $\mu$ is
recoverable from the Knizhnik-Polyakov-Zamolodchikov (KPZ) scaling
argument \cite{Knizhnik:1988ak}, we shall set $\mu = 1$ (in a
suitable unit) unless stated otherwise and avoid unnecessary
complication. In the spacetime description, this originates from
freedom of shifting the $\phi$ coordinate, and our convention is
equivalent to setting $\frac{\tau_\pp V_\pp}{E} = 2$.

The non-BPS hairpin brane\footnote{This non-BPS hairpin brane has
non-vanishing RR charge and was called (would-be) BPS in
\cite{Kutasov:2004dj}, so should be distinguished from the non-BPS
brane studied in \cite{Kluson:2004xc, Kluson:2004yk}.} in the
``throat geometry" is then given by direct product of noncompact
class-2 brane of the $\cN=2$ Liouville theory and appropriate Cardy
boundary state of the $SU(2)_{N-2}$ sector.\footnote{This is a
typical example of a D-brane in the noncompact Gepner model. For a
thorough study on this subject, see \cite{Eguchi:2004ik}.} The
precise form of the latter is not important and we take, for
instance, $|L=0\rangle$ corresponding to a D0-brane located at the
North pole on $\mathbb{S}^3$.

D-brane's shape is determined by the Cardy boundary states of the
$\cN=2$ Liouville theory. For a non-BPS hairpin brane, we take the
class-2 boundary states \cite{Eguchi:2003ik,Ahn:2003tt}\footnote{See
also
\cite{McGreevy:2003dn,Ribault:2003ss,Israel:2004jt,Fotopoulos:2004ut}
for related studies.}:
\bea | P,Q \rangle^{(\sigma)} = \int_{0}^\infty \dr p
\int_{-\infty}^{\infty} \dr q ~\Psi^{(\sigma)}_{P,Q}(p,q)
|p,q\rangle\rangle^{(\sigma)} \ , \nonumber \eea
labelled by $(P,Q)$. Here, $\sigma$ denotes the spin structure and
the Ishibashi states $\dket{p,q}^{(\sigma)}$ are defined by the
irreducible $\cN=2$ massive characters:
\begin{eqnarray}
\ch{(\sigma)}{}(p,q;\tau,z) = e^{2\pi i \tau (\frac{p^2}{2} +
\frac{q^2}{2})}
 e^{2\pi i \cQ q z} \frac{\th_{\lb \sigma \rb}(\tau,z)}{\eta(\tau)^3}\ .\nonumber
\end{eqnarray}
Wave functions of the class-2 boundary states are then given by
\bea \Psi_{P,Q}^{(\sigma)}(p,q) = \sqrt{2}\cQ e^{2\pi i \frac{Q
q}{\cQ}} \cos(2\pi Pp) \frac{\Gamma\left(i\cQ p\right)
\Gamma\left(1+i\frac{2p}{\cQ}\right)}
{\Gamma\left(\frac{1}{2}+i\frac{p}{\cQ} +\frac{q}{\cQ}
-\frac{\nu(\sigma)}{2}\right) \Gamma\left(\frac{1}{2}+i\frac{p}{\cQ}
-\frac{q}{\cQ}+ \frac{\nu(\sigma)}{2}\right)} \, , \label{hairpin
bw} \eea
where $\nu(\NS)=0$ and $\nu(\R)=1$. The parameter $Q$ appears only
through irrelevant phase-factor, so the class-2 boundary states are
classified solely in terms of the parameter $P$. The simplest
wave function, Eq.(\ref{hairpin bw}) with $P=0$, is the one
localized along the hairpin curve: $\exp(-{1 \over 2} {\cQ\phi}) =
2\cos\frac{\cQ x}{2}$.

As proposed in \cite{Nakayama:2004yx}, radion-rolling boundary
states are obtained by Wick rotation of the hairpin D-brane boundary
states. An important point is that the Wick-rotated momentum space
wavefunction is {\it not} just given by naive replacement $q \to
i\omega$ of Eq.(\ref{hairpin bw}) but also contains nontrivial
damping factor arising from noncompactness of the rolling D-brane
trajectory and nontrivial choice of the configuratoin space
integration contour. Applying the prescribed Wick rotation, the
rolling-radion boundary state is obtained as
\bea |B \rangle = \tau_p \int_0^\infty \dr p \int_{-\infty}^{\infty}
\dr \omega ~\Psi^{\sNS} (p, \omega) |p, \omega \rangle\rangle
\nonumber \eea
where the boundary wave function $\Psi(p, \omega)$ is given by
\bea \Psi^{\sNS}(p,\omega) &=& \frac{i
\sqrt{2}\cQ\sinh(\frac{ 2 \pi p}{\cQ})}
{2\cosh[\frac{\pi}{\cQ}(p+\omega)]\cosh[\frac{\pi}{\cQ}(p-\omega)]}
\cdot \frac{\Gamma(i\cQ p)\Gamma(1+i\frac{2p}{\cQ})}
{\Gamma(\frac{1}{2}-i\frac{\omega}{\cQ}+i\frac{p}{\cQ})
\Gamma(\frac{1}{2}+i\frac{\omega}{\cQ}+i\frac{p}{\cQ})} \nonumber \\
&=&
\frac{1}{\cQ}\frac{\sqrt{2}}{\pi}\frac{\Gamma(\frac{1}{2}+i\frac{\omega}{\cQ}
-i\frac{p}{\cQ})\Gamma(\frac{1}{2}-i\frac{\omega}{\cQ}-i\frac{p}{\cQ})
\Gamma(1+i\cQ p)}{\Gamma(1-\frac{2 i p}{\cQ})}  \ . \label{rob} \eea
This wave function defines stringy counterpart to Dp-brane's
trajectory in the supergravity approximation limit, $N = 2/{\cal
Q}^2 \rightarrow \infty$ \footnote{Note that this limit corresponds
to strong coupling limit of ${\cal N}=2$ Liouville theory. On the other hand, in the
T-dual, $SL(2,\mathbb{R})/U(1)$ theory, the limit corresponds to
weak coupling limit. The latter is therefore the correct description
of NS5-brane "throat geometry" in the supergravity approximation.}.
Indeed, in the latter limit, Fourier-transform of the wave function
Eq.(\ref{rob}) is seen peaked along the trajectory:
\bea e^{-\frac{1}{2} \cQ\phi} = 2\cosh\frac{\cQ t}{2} \  . \nonumber
\eea
This trajectory describes rolling radion and is in full accord with
the effective field theory result, Eq.(\ref{traj}).

Recall that, in constructing tachyon rolling boundary states, two
different approaches have been proposed. One approach is Sen's
construction \cite{sen-RT,sen-TM} and the other one is the time-like
boundary Liouville (TBL) construction \cite{Gutperle:2003xf}. The
main difference between the two is that Sen's boundary state is
expanded by the Ishibashi states constructed from not only the pure
momentum states but also full (Euclidean) $c=1$ primaries. Our
construction of radion rolling boundary state is closer in structure
to the TBL approach in that all the Ishibashi states have
normalizable inner products. Still, it is distinct from TBL approach
in detail in that no analytic continuation of CFT parameter (such as
$b \to i$ in TBL) is needed. We emphasize that this does not
necessarily imply that coupling to imaginary momentum modes $e^{-m
T}$ is zero. The spirit of the Liouville theory asserts that, on the
contrary, one can extract these one-point functions from analytic
continuation of the real momentum Ishibashi state expansion.\footnote{For related discussions on this point, see
\cite{deBoer:2003hd,Balasubramanian:2004fz,Fredenhagen:2004cj,Gaberdiel:2004na}.}
\subsect{Electrified D-brane rolling in ``throat geometry"}

Consider now an electrified Dp-brane rolling down the "throat
geometry" of NS5-branes. Within the effective field theory approach
studied in section \ref{2}, we found that the classical trajectory
of electrified Dp-brane is exactly the same as that of bare
Dp-brane except ``time dilation effect" $t \to
\sqrt{1-\varepsilon^2} t$. It indicates that the two might be
related by a Lorentz boost.

Indeed, in constructing radion rolling boundary state for an
electrified Dp-brane, intuitive and technically simple way is to
utilize stringy version of the Lorentz boost. Such prescription was
already worked out in \cite{Rey:2003zj} and involves successive
application of T-duality, Lorentz boost and inverse T-duality. In
\cite{Rey:2003zj}, the prescription was mainly applied to tachyon
rolling of an electrified unstable Dp-brane, but the prescription
is general enough and hence is equally applicable to the radion
rolling of Dp-brane. In practice, the prescription can be
summarized as taking the following replacement to the boundary state
of a bare Dp-brane:
\bea |0\rangle &\to& \gamma^{-1}|0\rangle
\qquad\qquad t \to \gamma^{-1} t \qquad\qquad
\omega \to \gamma \omega \nonumber \\
\nonumber \\
 \left(
  \begin{array}{c}
    \alpha^0 \\
    \alpha^1 \\
  \end{array}
  \right) &\to& \Lambda^{-1} \left(
  \begin{array}{c}
    \alpha^0 \\
    \alpha^1 \\
  \end{array}
  \right) \qquad\qquad\qquad\quad \left(
  \begin{array}{c}
    \psi^0 \\
    \psi^1 \\
  \end{array}
  \right) \to \Lambda^{-1} \left(
  \begin{array}{c}
    \psi^0 \\
    \psi^1 \\
  \end{array}
  \right) \nonumber  \\
   \left(
  \begin{array}{c}
    \overline{\alpha}^0 \\
    \overline{\alpha}^1 \\
  \end{array}
  \right) &\to& \,\, \Lambda \, \left(
  \begin{array}{c}
    \overline{\alpha}^0 \\
    \overline{\alpha}^1 \\
  \end{array}
  \right)  \qquad\qquad\qquad\qquad
 \left(
  \begin{array}{c}
    \overline{\psi}^0 \\
    \overline{\psi}^1 \\
  \end{array}
  \right) \to \,\, \Lambda \, \left(
  \begin{array}{c}
    \overline{\psi}^0 \\
    \overline{\psi}^1 \\
  \end{array}
  \right)  \label{RSrep}
  \eea
  where
  \bea
  \Lambda =  \gamma \left( \begin{array}{cc} 1 & +\varepsilon \\
  +\varepsilon & 1 \end{array} \right)
  \qquad\quad
   \Lambda^{-1} =  \gamma \left( \begin{array}{cc} 1 & -\varepsilon \\
   -\varepsilon & 1 \end{array} \right)
 \qquad\quad \gamma \equiv \frac{1}{\sqrt{1 - \varepsilon^2}}\, , \nonumber
\eea
and $\alpha$, $\psi$ denote bosonic and fermionic free field
oscillators. Using the prescription, we shall now construct the
radion rolling boundary state of an electrified Dp-brane and show
that they yield in the supergravity limit to classical trajectory
and conserved currents that agree with those obtained in section
\ref{2} from the effective field theory approach.

In applying the prescription Eq.(\ref{RSrep}), one main obstacle is
that the starting theory --- Lorentzian $\cN=2$ Liouville theory
$(\phi,T=X^0)$ coupled to free $X^1$ superconformal field theory ---
is an interacting conformal field theory. In particular, the
boundary state of the former is spanned by the $\cN=2$ Virasoro
module and not by the Fock module on which the replacement
Eq.(\ref{RSrep}) is readily definable. For example, the A-type
Ishibashi boundary state is expanded as:
\begin{equation}
\left|A\rangle\rangle \right. =\Big(\cdots +
\f{i\sigma}{2h_0-q_0}G^+_{-\f{1}{2}}\overline{ G}^+_{-\f{1}{2}}
+\f{i\sigma}{2h_0+q_0}G^-_{-\f{1}{2}}\overline{
G}^-_{-\f{1}{2}}+1\Big) \left|j,q\rangle \right. ,\label{Ishi}
\end{equation}
where $\sigma = \pm1$ is the spin structure, $2h_0 = -j(\cQ+j) +
q^2$ and $q_0 = -q\cQ$.\footnote{Roughly speaking, $|j,q\rangle =
e^{j\phi + iqX} |0\rangle =
e^{(-\frac{\cQ}{2}+ip)\phi+iqX}|0\rangle$.} Utilization of the
Virasoro module (instead of the Fock module) has been an important
step in solving the boundary Liouville theory via the so-called
modular bootstrap method \cite{Zamolodchikov:2001ah,Eguchi:2003ik}.
Determination and classification of possible boundary wave functions
in the Liouville theory crucially relied on such features.

We believe this is a technical issue the physics itself gets around.
Operationally, one can drop out the Liouville potential and realize
the $\cN =2$ Virasoro algebra by using the free field Fock
representation Eq.(\ref{SCA-L}). This is certainly justifiable in
the weak coupling end, $\phi \to +\infty$, in the "throat geometry",
where the closed string modes evolve freely. With such proviso, at
level-$1/2$, we may replace in Eq.(\ref{Ishi})
\bea G^+_{-\f{1}{2}} ~\rightarrow~ -{(q-j)}\Psi^+_{-\f{1}{2}};\quad
\qquad G^-_{-\f{1}{2}}~\rightarrow~ -{(q+j)}\Psi^-_{-\f{1}{2}}
\nonumber \eea
and obtain
\bea |A\rangle\rangle =\left(\cdots + i\sigma \f{(q-j)}{(q+j+\cQ)}
\Psi^+_{-\f{1}{2}}\overline{\Psi}^+_{-\f{1}{2}} +i\sigma
\f{(q+j)}{(q-j-\cQ)}\Psi^-_{-\f{1}{2}}\overline{\Psi}^-_{-\f{1}{2}}+1\right)
|j,q\rangle \ . \label{foc} \eea
We can thus formally rewrite the $\cN=2$ Virasoro Ishibashi states
in terms of the $\cN=2$ free-field Fock module. Having expressed
rolling-radion boundary states for a bare D-brane in terms of the
Fock module, we can utilize the prescription Eq.(\ref{RSrep}) and
construct rolling-radion boundary states for an electrified D-brane.
Actually, as we will see in the next section, Fock module
realization is quite imperative not only for extracting wave
functions and classical trajectory therein but also for extracting
conserved currents.

Applying the prescription Eq.(\ref{RSrep}) to the bare D-brane
boundary states Eq.(\ref{rob}), we finally obtain the electrified
rolling-radion boundary state
\bea | B, \varepsilon \rangle = \tau_p \int_{0}^\infty \dr p
\int_{-\infty}^{\infty} \dr \omega ~\Psi_\varepsilon (p,\omega) |
p,\omega;\varepsilon \rangle\rangle \ , \nonumber \eea
where the modified Ishibashi states $ | p,\omega;
\varepsilon\rangle\rangle \ $ is constructed accordingly from the
prescription Eq.(\ref{RSrep}). The wave function
$\Psi_\varepsilon(p, \omega)$ is then extracted as
\bea \Psi_\varepsilon (p,\omega) =
\frac{i \sqrt{2}\cQ \sinh(\frac{ 2 \pi p}{\cQ})}
{2\cosh[\frac{\pi}{\cQ}(p+\gamma\omega)]\cosh[\frac{\pi}{\cQ}(p-\gamma\omega)]}
\cdot \frac{\Gamma(i\cQ p)\Gamma(1+i\frac{2p}{\cQ})}
{\Gamma(\frac{1}{2}-i\frac{\gamma\omega}{\cQ}+i\frac{p}{\cQ})
\Gamma(\frac{1}{2}+i\frac{\gamma\omega}{\cQ}+i\frac{p}{\cQ})} \ .
\label{0mode} \eea

{} From the zero mode part Eq.(\ref{0mode}) alone, we can draw variety of
physics intuition\footnote{Aspects of the higher-level parts will be
discussed in the context of conserved currents in the next
section.}. For example, the zero-mode part yields information on
radion-rolling trajectory of the electrified D-brane. By comparing
Eq.(\ref{0mode}) with the first expression of Eq.(\ref{rob}), we
immediately find that the position space wave function of
electrified D-brane is given by $\Psi_\varepsilon (\phi,t) =
\Psi_0(\phi,\sqrt{1-\varepsilon^2}t)$, where $\Psi_0(\phi,t)$ is the
position space wave function of bare D-brane obtained by Fourier
transform of Eq.(\ref{rob}). As emphasized in the previous
subsection, in the supergravity limit $N \rightarrow \infty$,
$\Psi_0(\phi,t)$ is localized on the radion rolling trajectory of
the bare D-brane, $\Psi_0(\phi,t) \sim \delta (\phi - \phi_0(t))$.
Thus, our proposed boundary wavefunction is localized in the rolling
trajectory of electrified D-brane with expected ``Lorentz time
dilation" $t\to \gamma^{-1} t$:
\bea \Psi_\varepsilon (\phi,t) =
\Psi_0(\phi,\sqrt{1-\varepsilon^2}t) \sim \tau_p \delta (\phi -
\phi_0(\gamma^{-1}t)) \ . \nonumber \eea
This trajectory agrees with the one obtained from the effective
field theory analysis in section \ref{2}.

Comments are in order.
\begin{itemize}
    \item The boundary interaction which represents the class-2 brane
    in the (Euclidean) $\cN = 2$ Liouville theory was studied in
    \cite{Ahn:2003wy,Ahn:2004qb,Hosomichi:2004ph}. In order to accommodate the constant
electric field background, one needs to modify the boundary
interaction accordingly. Our proposal is that the prescription
Eq.(\ref{RSrep}) to the boundary interaction is the correct one
(after the appropriate Wick rotation). To all orders in worldsheet
perturbation theory, the resulting boundary interaction is marginal.
    \item By construction, it is not apparent that the prescription Eq.(\ref{RSrep})
    is compatible with the $\cN=2$ worldsheet superconformal symmetry.
    On the other hand, the $\cN=1$ part of it, which is to be gauged,
    is manifestly preserved. To show this, consider a boundary state $|B \rangle$ obeying
    $\cN=1$ superconformal boundary condition and a new boundary state
    ${\cal P}_0 \mathbb{U}_\varepsilon |B \rangle$, where ${\cal P}$ stands for the projection
    operator onto zero-winding modes and $\mathbb{U}_\varepsilon$ refers to the
    left-right asymmetric Lorentz boost by $\varepsilon$. Notice
    that the combined operation ${\cal P}_0 \mathbb{U}_\varepsilon$ is
    precisely the prescription Eq.(\ref{RSrep}). We now prove that
    the boosted boundary state ${\cal P}_0 \mathbb{U}_\varepsilon |B
    \rangle$ obeys the $\cN=1$ conformal boundary condition as well:
\begin{eqnarray}
(G_{r} + i\sigma \overline{G}_{-r}) {\cal P}_0
\mathbb{U}_\varepsilon |B\rangle =0 \ . \nonumber
\end{eqnarray}
Since ${\cal P}_0$ commutes with $(G_{r} + i\sigma
\overline{G}_{-r})$ and since the $\cN=1$ boundary condition, in
contrast to the full $\cN=2$ superconformal symmetry, is invariant
under the Lorentz boost along the $\mathbb{R}^{1,5}$ plane
(longitudinal to NS5-brane), it follows that
\bea (G_{r} + i\sigma \overline{G}_{-r}) {\cal P}_0
\mathbb{U}_\varepsilon |B\rangle &=&
{\cal P}_0 \mathbb{U}_\varepsilon \mathbb{U}_\varepsilon^{-1}
(G_{r} + i\sigma \overline{G}_{-r})\mathbb{U}_\varepsilon |B\rangle \nonumber \\
&=&  {\cal P}_0 \mathbb{U}_\varepsilon (G_{r} + i\sigma \overline{G}_{-r})|B\rangle \nonumber
\\
&=& 0 \ , \nonumber \eea
where in the last line, we have used the property that the original
boundary states $|B\rangle$ satisfy the $\cN=1$ conformal boundary
condition. Similar consideration for the worldsheet energy-momentum
tensor leads to the same conclusion. We thus see that any boundary
states obtained by the prescription Eq.(\ref{RSrep}) are indeed
consistent boundary states.
\end{itemize}

\sect{Conserved Currents of Rolling Radion Boundary States}\label{4}
In understanding radion-rollong dynamics, conserved charges provide
indispensible information. So, having obtained the exact boundary states,
we shall now investigate such conserved charges. In section \ref{2},
the conserved charges were computed in the effective field theory
approach. We shall see that, in the supergravity limit $N = {2 \over {\cal Q}^2} \rightarrow \infty$, conserved
charges extracted from exact boundary states agree with those from
the effective field theory.

Again, we shall focus on D-brane dynamics at weak coupling region,
where the ${\cal N}=2$ Liouville potential is dropped off and the
"throat geometry" is describable by ${\cal N}=2$ linear dilaton
conformal field theory. Expanding in terms of free Fock modules, the
boundary state in the linear dilaton background $\Phi(x) = V_\mu
x^\mu$ takes the form
\bea |B\rangle = \tau_p \int \f{ \dr^d {\bf k}}{(2\pi)^d}
\Big[\widetilde{B}({\bf k})-i\sigma\widetilde{A}_{\mu\nu}({\bf k})
\psi_{-\frac{1}{2}}^{\mu}\overline{\psi}_{-\frac{1}{2}}^\nu+\cdots\Big]
|{\bf k}\rangle_0\ , \qquad {\bf k}_\mu \in \mathbb{R}^d \ .
\nonumber \eea
Energy-momentum tensor of the boundary state is then given by (see
{\it e.g.} \cite{Karczmarek:2003xm})
\bea T_{\mu\nu}({x})=-e^{-V_\mu x^\mu} \Big(
A_{\mu\nu}({x})+B({x})\eta_{\mu\nu} \Big) \ . \label{EMtensor B}
\eea
in the convention of the Fourier transformation: \bea f({x})=\int
\f{\dr^D {k}}{(2\pi)^D} e^{i{k_\mu}{x^\mu}}\widetilde{f}({k})\ .
\nonumber \eea
It is readily seen that the energy-momentum tensor Eq.(\ref{EMtensor
B}) satisfies the shifted conservation laws:
\bea
\partial^\mu T_{\mu\nu}=\f{1}{2}V_\nu U\ ,\quad \frac{1}{4}U =
e^{-{V_\mu}{x^\mu}} B \ .
\label{conse}
\nonumber \eea
\subsection{Conserved currents for bare D-brane}
We shall first examine conserved currents for the bare D-brane case
and compare them with the effective field theory results. Expand the
boundary state in free field Fock states:
\bea |B\rangle = \tau_\pp \Big[ B -i\sigma (A_{(\mu\nu)}+C_{[\mu\nu]})
\psi_{-\frac{1}{2}}^\mu \overline{\psi}_{-\frac{1}{2}}^\nu+\cdots
\Big] |0\rangle  \ , \label{thisbdry} \eea
where $A_{\mu\nu}$, $C_{\mu\nu}$ refer to symmetric and
antisymmetric irreducible parts. With the linear dilaton background
along the Liouville direction, the energy-momentum tensor is given
by
\bea T_{\mu\nu} = - \tau_\pp~e^{\frac{1}{2} \cQ\phi}(A_{\mu\nu} +
\eta_{\mu\nu}~B) \ , \nonumber \eea
while the string current tensor is given by
\bea Q_{\mu\nu} = -\tau_\pp ~ e^{\frac{1}{2} \cQ\phi}C_{\mu\nu} \ .
\nonumber \eea
We expect them to reduce to the effective field theory results of
section \ref{2} in the supergravity limit, $N ={2 \over {\cal Q}^2}
\rightarrow \infty$.

{}The first term $B$ in Eq.(\ref{thisbdry}) involves no Fock
oscillator, so it is extractible from the boundary wave function
Eq.(\ref{rob}) by making Fourier transform. A useful integral
formula is
\begin{eqnarray}
\frac{1}{\cQ}\int_{-\infty}^{\infty} \frac{\dr
\omega}{2\pi}e^{-i\omega
t}\frac{\Gamma(\frac{1}{2}+i\frac{\omega}{\cQ}-i\frac{p}{\cQ})
\Gamma(\frac{1}{2}-i\frac{\omega}{\cQ}-i\frac{p}{\cQ})}{\Gamma(1-\frac{2
i p}{\cQ})} =
\left[2\cosh\left(\frac{t\cQ}{2}\right)\right]^{\frac{2ip}{\cQ}-1} \
. \label{formula1}
\end{eqnarray}
In the supergravity limit $N = {2 \over {\cal Q}^2} \rightarrow
\infty$, the $\Gamma(1+i\cQ p)$ factor in Eq.(\ref{rob}) may be
dropped off. The result is
\bea B(\phi,t) &=& \frac{1}{\cQ}\int_{-\infty}^{\infty} \frac{\dr
\omega \dr p}{(2\pi)^2} e^{ip\phi-i\omega t}
\frac{\sqrt{2}}{\pi}\frac{\Gamma(\frac{1}{2}+i\frac{\omega}
{\cQ}-i\frac{p}{\cQ})\Gamma(\frac{1}{2}-i\frac{\omega}{\cQ}-i\frac{p}{\cQ})}
{\Gamma(1-\frac{2 i p}{\cQ})} \nonumber \\
&=& \frac{1}{\cosh(\frac{t\cQ}{2})} \frac{\sqrt{2}}{2\pi}\delta
\Big(\phi+ \frac{2}{\cQ} \log(2\cosh\frac{\cQ t}{2}) \Big) \nonumber \\
&\equiv& \frac{\sqrt{2}}{2\pi} B_0(\phi, t) \ . \label{b0} \eea
The numerical factor $\frac{\sqrt{2}}{2\pi}$ is absorbable by
rescaling the string coupling constant $g_{\rm st}$. The
$(00)$-component of the second term $A_{(\mu\nu)}$ in
Eq.(\ref{thisbdry}) is obtainable from Eq.(\ref{foc}) and
Eq.(\ref{rob}). We first Wick-rotate, expand the free fermion
oscillators into components
\bea \Psi^{\pm}\overline{\Psi}^{\pm} =
\frac{1}{2}\left(\psi^X\overline{\psi}^X\pm
i\psi^{\phi}\overline{\psi}^X \pm i\psi^X\overline{\psi}^\phi -
\psi^\phi\overline{\psi}^\phi \right) \ , \nonumber \eea
and extract the
$\psi^0_{-\frac{1}{2}}\overline{\psi}^0_{-\frac{1}{2}}$ component
from the boundary states. In the supergravity limit $N ={2 \over
{\cal Q}^2} \rightarrow \infty$, the factor $\Gamma(1 + i {\cal Q}
p)$ can be dropped off. With Fock module realization $j \rightarrow
i p - {\cal Q}/2$ and Wick rotation $q \rightarrow i \omega$, we
find that
\bea \frac{2\pi}{\sqrt{2}}\widetilde{A}_{00}(p,\omega) &=& -(-i)^2{1
\over {\cal Q}} \frac{\Gamma(\frac{1}{2}
+i\frac{\omega}{\cQ}-i\frac{p}{\cQ})\Gamma(\frac{1}{2}
-i\frac{\omega}{\cQ}-i\frac{p}{\cQ})}{\Gamma(1-\frac{2 i p}{\cQ})}
\left(\frac{i\omega -ip+\frac{\cQ}{2}}{i\omega + ip+\frac{\cQ}{2}}
+ \frac{i\omega + ip-\frac{\cQ}{2}}{i\omega-ip-\frac{\cQ}{2}}\right) \nonumber \\
 &=& - {1 \over {\cal Q}} \frac{\Gamma(\frac{1}{2}+1+i\frac{\omega}{\cQ}-i\frac{p}{\cQ})
 \Gamma(\frac{1}{2}-1-i\frac{\omega}{\cQ}-i\frac{p}{\cQ})}{\Gamma(1-\frac{2 i p}{\cQ})}
 \nonumber \\
 &&- {1 \over {\cal Q}}
 \frac{\Gamma(\frac{1}{2}-1+i\frac{\omega}{\cQ}-i\frac{p}{\cQ})\Gamma(\frac{1}{2}+1
 -i\frac{\omega}{\cQ}-i\frac{p}{\cQ})}{\Gamma(1-\frac{2i p}{\cQ})} \nonumber \\
 &\equiv& \widetilde{A}_0 (p, \omega)
\ . \nonumber \eea
In the first line, the first sign factor $(-)$ is due to relative
sign convention between Eq.(\ref{foc}) and Eq.(\ref{thisbdry}),
while the second sign factor $(-i)^2$ arises from the Wick rotation
from $A_{XX}$ to $A_{00}$. Fourier transforming the amplitude by
formally substituting $\omega \to \omega \pm i\cQ$ into each term of
Eq.(\ref{formula1}) and modifying the integration
contour\footnote{This is a formal manipulation since the direct
calculation yields a divergent result. We should regard it as an
analytic continuation or a suitable Wick rotation from the
corresponding Euclidean calculation, where no divergence arises. See
\cite{Nakayama:2004yx} on this issue.}, we find
\bea A_{00}(\phi,t) &=& \int_{-\infty}^{\infty} \frac{\dr \omega \dr
p}{(2\pi)^2} e^{ip\phi-i\omega t}
\widetilde{A}_{00}(p,\omega) \nonumber \\
&=& - \frac{\cosh(t\cQ)}{\cosh (\frac{t\cQ}{2})}
\frac{\sqrt{2}}{2\pi}\delta\Big(\phi+{2 \over {\cal Q}} \log (2
\cosh {{\cal Q} t \over 2}) \Big) \nonumber \\
&\equiv& {\sqrt{2} \over 2 \pi} A_0(\phi, t) \ . \label{a0} \eea
{} We also find that $A_{11}(\phi, t)$ equals to the zero-mode wave
function $B(\phi, t)$ since it comes from the flat Dirichlet
Ishibashi states. Then, independent components of the
energy-momentum tensor are\footnote{All other components are
obtained from the conservation law of the energy momentum tensor. In
the boundary conformal field theory language, this is equivalent to
the statement that the boundary states preserve (half of) the
$\cN=1$ superconformal symmetry.}
\bea T_{00}(\phi, t) &=& -\tau_\pp~e^{{1 \over 2} {\cal Q} \phi}
(A_{00} - B) =
\delta (\phi - \phi_0(t)) \nonumber \\
T_{ij}(\phi, t) &=& - \tau_\pp~e^{{1 \over 2} {\cQ} \phi} (A_{11} + B)
= - {\rm sech}^2\Big({{\cal Q} t \over 2} \Big)
\delta_{ij}\delta(\phi - \phi_0(t))  \nonumber \eea
where $\phi_0(t)$ is the trajectory in the supergravity limit
\bea \phi_0(t) = - {2 \over {\cal Q}} \log (2 \cosh {{\cal Q} t
\over 2})~. \label{classical} \eea
The result is in agreement with the result obtained from effective
field theory approach in section 2.2.\footnote{Recall that we have
chosen shift of the $\phi$ coordinate so that $\tau_p V_p /E = 2$.}

Of particular interest is exploring stringy effects present at
finite $N$ and $\cQ$. Repeating the analysis but retaining the
factor $\Gamma(1 + i {\cal Q}p)$ in the boundary wave function, we
now find that
\bea B(\phi,t) = \f{\sqrt{2}} {\pi{\cQ}\Big(2 \cosh \f{\cQ
t}{2}\Big)^{\frac{2}{\cQ^2}+1}}
\exp\left[-\f{\phi}{\cQ}-\f{e^{-\frac{\phi}{\cQ}}} {\Big(2 \cosh
\f{\cQ t}{2}\Big)^{\f{2}{\cQ^2}}}\right] \ . \label{bxt} \eea
We thus learn that the classical trajectory is now smeared to the
profile:
\bea \f{1}{\cQ}\f{1} {\Big(2 \cosh \f{\cQ
t}{2}\Big)^{\frac{2}{\cQ^2}}}
\exp\left[-\f{\phi}{\cQ}-\f{e^{-\frac{\phi}{\cQ}}} {\Big(2 \cosh
\f{\cQ t}{2}\Big)^{\f{2}{\cQ^2}}}\right] \ . \nonumber \eea
In turn, the energy-momentum tensor at finite $N ={2 \over {\cal
Q}^2}$ is given by
\bea T_{00}(\phi, t) =
{1 \over {\cal Q}}
~\left(\f{e^{- {{\cal Q} \phi \over 2}}} {2 \cosh \f{\cQ
t}{2}}\right)^{\frac{2}{\cQ^2}-1} \cdot \exp\left[-\left(
\f{e^{-\frac{\cQ \phi}{2}}} {2 \cosh \f{\cQ t}{2}}
\right)^{\f{2}{\cQ^2}}\right] \ , \label{stringy} \eea
replacing the delta function localized energy density at
supergravity limit into smeared distribution in $\phi$-direction.
One can check readily that, after integrating over $\phi$, the total
energy is a constant of motion. The results are in full accord with
expectation that stringy effects would smear out the classical
trajectory to a distribution of height and width set by $ {\cal Q} =
\sqrt{2/N}$.

To gain intuition how the stringy effects smear out the trajectory,
consider the energy density $T_{00}(\phi, t)$ and examine its
behavior both near asymptotic past and future $t \rightarrow \pm
\infty$ and around time-symmetric turning point $t \sim 0$. For the
trajectory in the supergravity limit, $\phi(t)$ evolves as
\bea \phi_0(t) \simeq \left\{ \begin{array}{ccc} -|t| \mp {2 \over
{\cal Q}} e^{-|t|{\cal Q}} + \cdots  & \qquad\mbox{for} \qquad &
t \gg \pm {2 \over {\cal Q}} \\
&&\\
 -{2 \over {\cal Q}} \log{ 2} + {{\cal Q} \over 4} |t|^2 +
\cdots & \qquad \mbox{for} \qquad & |t| \sim 0 \end{array} \right. \
. \nonumber \eea
For the stringy profile at finite $N = {2 \over {\cal Q}^2}$,
Eq.(\ref{stringy}) clearly reveals that the profile is a
Poisson-type distribution. Maximum of the energy density
distribution is now located at
\bea \f{e^{- {{\cal Q} \phi \over 2}}} {2 \cosh \f{\cQ t}{2}} = 1 -
{{\cal Q}^2 \over 2} \ , \nonumber \eea
so it deviates from the trajectory $\phi_0(t)$ in the supergravity
limit by $\Delta \phi \equiv (\phi(t) - \phi_0(t)) \simeq {\cal Q}$.
In fact, for $\phi = \phi_0(t)$, each factor in round bracket in
Eq.(\ref{stringy}) is reduced to unity. The variance $\Delta \phi$
of the distribution may be estimated more accurately by expanding
$\phi(t)$ around $\phi_0(t)$ in each factor of round brackets in
Eq.(\ref{stringy}). We find that
\bea T_{00}(\phi, t) \simeq T_{00} \Big|_{\rm max} \Big(1 - {1 \over
2} N (N-1) (\Delta \phi)^2 + \cdots \Big)\ , \nonumber \eea
and hence estimate the variance of the energy-density distribution
as
\bea \Delta \phi \simeq \sqrt{2 {\cal Q}^2 \over 2 - {\cal Q}^2}\ .
\label{variance} \eea
The result bears two interesting implications. First, unlike
rolling-tachyon boundary states for unstable D-brane for which the
boundary state exhibits a universal variance of the boundary wave
function set by the string scale $\sqrt{\alpha'}$, rolling-radion
boundary state exhibits ${\cal Q}$-dependent variance. We interpret
this as a feature originating from the fact that D-brane is now
placed in curved background of "throat geometry" rather than flat
Minkowski spacetime, as considered for unstable D-brane. Second, both
the energy density Eq.(\ref{stringy}) and the variance
Eq.(\ref{variance}) indicate that $\cQ^2 = 2$ is a special point.
Notice, however, that this highly stringy regime does not correspond
to a NS5-brane background. Rather, being $N=1$, this describes a
conifold background. Indeed, it has been recurrently noted that
closed string sector of  ${\cal N}=2$ Liouville theory,
equivalently, ${\cal N}=2$ $SL(2,\mathbb{R})/U(1)$ theory changes
drastically across this selfdual point $\cQ^2 = 2$. This phenomenon has to do
with the fact that the chiral ${\cal N}=2$ Liouville potential
violates the Seiberg bound. Our observation is that open string
sector of these theories also changes drastically across the same
selfdual point. In the next section, we shall also discover another
manifestation of such effect in the context of closed string
radiation out of radion rolling.

\subsection{Conserved currents for electrified D-brane}
We shall now extend analysis of conserved currents to the
electrified D-brane. After turning on boundary interaction for the
electric field, the rolling-radion boundary state is modified
according to the prescription Eq.(\ref{RSrep}).
One readily finds that the prescription Eq.(\ref{RSrep}) gives rise
to the following transformation to the boundary states:
\bea
B(\phi, t) &=& + \gamma^{-1}~ B_0(\phi,\gamma^{-1}t) \nonumber \\
A_{00}(\phi, t) &=& + \gamma~ A_0(\phi,\gamma^{-1}t) - \varepsilon^2
~\gamma B_0(\phi,\gamma^{-1}t)
\nonumber \\
C_{01}(\phi, t) &=& + \gamma~ \varepsilon~ (A_0(\phi,\gamma^{-1}t)-
B_0~(\phi,\gamma^{-1}t))
\nonumber \\
A_{11}(\phi, t) &=& +\gamma ~B_0(\phi,\gamma^{-1}t) -
\varepsilon^2~\gamma A_{0}
(\phi,\gamma^{-1}t) \nonumber \\
A_{ij}(\phi, t) &=& +\gamma^{-1}~ B_0(\phi,\gamma^{-1}t)~\delta_{ij}
\ , \nonumber \eea
where now $i,j$ run for space transverse to the boost direction.
Here, $A_0(\phi,t) $ and $B_0(\phi,t)$ are components of the
boundary wave functions for bare D-brane given in Eq.(\ref{a0}) and
Eq.(\ref{b0}), respectively. In the supergravity limit $N
\rightarrow \infty$ and ${\cal Q} \rightarrow 0$, using $B_0, A_0$
obtained in the previous subsection, it is straightforward to check
that the boundary wavefunction is reduced to the result of effective
field theory analysis in section 2.2.

By construction, the energy-momentum tensor transforms covariantly
under the Lorentz boost and hence is always conserved. Still,
it would be illuminating to check this by direct computations. So,
consider $T_{0\phi} = e^{\frac{1}{2} \cQ\phi} A_{0\phi}$, viz.
momentum density along $\phi$-direction. It suffices to examine it
in the supergravity limit $N = {2 \over {\cQ}^2} \rightarrow
\infty$:
\bea \frac{2\pi}{\sqrt{2}}\widetilde{A}_{0\phi}(p,\omega) &=& {1
\over {\cal Q}}
\frac{\Gamma(\frac{1}{2}+i\frac{\gamma\omega}{\cQ}-i\frac{p}{\cQ})
\Gamma(\frac{1}{2}-i\frac{\gamma\omega}{\cQ}-i\frac{p}{\cQ})}{\Gamma(1-\frac{2i
p}{\cQ})} \left(\frac{i\gamma\omega -ip+\frac{\cQ}{2}}{i\gamma\omega
+ ip+\frac{\cQ}{2}} - \frac{i\gamma\omega +
ip-\frac{\cQ}{2}}{i\gamma\omega-ip-\frac{\cQ}{2}}\right)
\nonumber\\
 &=& - {1 \over {\cal Q}} \frac{\Gamma(1+\frac{1}{2}+i\frac{\gamma\omega}{\cQ}-i\frac{p}{\cQ})
 \Gamma(-1+\frac{1}{2}-i\frac{\gamma\omega}{\cQ}-i\frac{p}{\cQ})}
 {\Gamma(1-\frac{2i p}{\cQ})} \nonumber \\
 && +{1 \over {\cal Q}} \frac{\Gamma(-1+\frac{1}{2}+i\frac{\gamma\omega}{\cQ}-i\frac{p}{\cQ})
 \Gamma(1+\frac{1}{2}-i\frac{\gamma\omega}{\cQ}-i\frac{p}{\cQ})}
 {\Gamma(1-\frac{2i p}{\cQ})}\ . \nonumber
\eea
Fourier transforming back to $(\phi, t)$-space,
\bea A_{0\phi}(\phi,t) &=& \int_{-\infty}^{\infty} \frac{\dr \omega
\dr p}{(2\pi)^2} ~e^{ip\phi-i\omega t} ~A_{0\phi}(p,\omega)
\nonumber \\
&=& -\frac{\sinh(\gamma^{-1}\cQ t)}{\cosh (\frac{\gamma^{-1}\cQ
t}{2})} ~\frac{\sqrt{2}}{2\pi}\delta\Big(\phi + {2 \over {\cal Q}}
\log (\cosh {\gamma^{-1} {\cal Q} t \over 2} )\Big) \ , \nonumber
\eea
we find it agrees with the effective field theory result
Eq.(\ref{emtens}). It demonstrates that the prescription
Eq.(\ref{RSrep}) manifestly retains conservation of the
energy-momentum tensor (as well as other boundary state currents)
and BRST symmetry of the closed string equation of motion in the
presence of the boundary state as a source.

Again, it is of particular interest to explore stringy effects
present at finite $N$ and ${\cal Q}$ and to see how the
electrification intertwine with such stringy effects. Proceeding
analogously to the bare D-brane case, we find that the energy
density now reads
\bea T_{00}(\phi, t) = {\gamma \over \cQ} ~\left(
{e^{-\frac{\cQ\phi}{2} } \over 2 \cosh \frac{\cQ \gamma^{-1}t}{2} }
\right)^{ {2 \over \cQ^2} - 1} \cdot
\exp\left[-\left(\f{e^{-\frac{\cQ \phi}{2}}} {2 \cosh \f{\cQ
\gamma^{-1}t}{2}}\right)^{\f{2}{\cQ^2}}\right] \ . \nonumber \eea
This clearly indicates that the trajectory Eq.(\ref{traj}) in the
supergravity limit $N = {2 \over \cQ^2} \rightarrow \infty$ is now
replaced by the same Poisson-like distribution as for the bare
D-brane case. Note that, as expected, Lorentz covariance rendered
the energy density distribution get modified by overall $\gamma$
Lorentz boost and time dilation $t \rightarrow \gamma^{-1} t$. Thus,
electrifying the D-brane left no effect on the functional form of
the energy profile. Moreover, variance of the distribution $\Delta
\phi$ is Lorentz invariant, viz. the same for both the bare D-brane
and the electrified one. The value $\cQ^2 = 2$ is still the selfdual
critical point. This fact further supports the interpretation that
the selfdual critical point has intrinsically to do with closed
string sector of the ${\cal N}=2$ Liouville and
$SL(2,\mathbb{R})/U(1)$ theories.

It is straightforward to extract other conserved currents coupled to
higher oscillator modes out of the boundary states. As discussed in
\cite{Rey:2003zj}, the prescription Eq.(\ref{RSrep}) renders
components of conserved currents coupled to the oscillator along
electric field strength tensor directions are Lorentz boosted by
appropriate powers of the Lorentz factor $\gamma$. This effect then
serves as a useful mnemonic for understanding the boundary states
and higher conserved currents therein \cite{toappear}. It is also
possible to consider magnetifying D-brane by Lorentz rotation. Since
the procedure is also prescribed in \cite{Rey:2003zj} and is
sufficiently trivial, we shall not repeat it here.

\sect{Radiative Decay of Electrified D-Brane}\label{5}
Having constructed exact boundary states of an electrified D-brane,
we now investigate aspects of D-brane's radiative decay as it rolls
down the ``throat geometry" and eventually dissolve into the
NS5-brane. Final state of the process is a bound-state of the
electrified D-brane and the NS5-brane. Since the bound-state is of
non-threshold type, we expect that excess binding energy would be
released partly into closed string emission and, if present, partly
into some sort of ``radion matter".

Again, by the sequence of T-duality, Lorentz boost and inverse
T-duality, in an appropriate (de)compactification limit of the boost
direction, the radiative decay process of an electrified D-brane
would be related Lorentz covariantly to that of a bare D-brane. This
line of reasoning is elementary and intuitive, so we shall first
spell out the argument explicitly and then check it by direct
computations.

\subsect{Bare versus electrified: Lorentz boost and T-dual}
Consider a bare D$p$-brane rolling down the ``throat geometry".
Since the motion is rigid and homogeneous, we will compactify a
worldvolume direction common to the Dp-brane and the NS5-brane into
a circle of radius $R$. We shall then take $R \rightarrow 0$ limit.
T-dualizing along the circle direction, we now have D(p-1)-brane on
a dual circle of radius $2/R$, which is then decompactified in the
$R \rightarrow 0$ limit. Let us now boost the D(p-1)-brane and
measure the average number density of the closed strings emitted for
a fixed transverse mass $M$, $\langle N(M) \rangle/ V_{\rm p}$:
\bea {\langle N(M) \rangle \over V_{\rm p}} = \sum \hskip-0.5cm \int
\dr {\rm Vol} {1 \over 2 \omega} \vert \Psi(\omega) \vert^2 \ ,
\nonumber \eea
where the integral is over the phase-space of emitted quanta and
$\Psi(\omega)$ is the transition amplitude. Obviously, this is the
quantity independent of choice of the reference frame, so it ought
to be expressible as an integral over final state phase-space
variables either in the rest frame or in the boosted frame.

To demonstrate the Lorentz invariance of $\langle N(M) \rangle /
V_{\rm p}$, begin with the phase-space integral of the Dp-brane in
the rest frame. The on-shell energy is given by $E_p =
\sqrt{p^2 + {\bf k}^2 + M^2}$ where $p$ is the momentum along
$\phi$-direction, ${\bf k}$ is the momentun along (5-p) relative
transverse directions to the Dp-brane in $\mathbb{R}^5$ and $M$ is a
fixed transverse mass. Expressing it in terms of spectral variable,
the average number density for the NS-sector is given by
\bea {\langle N(M) \rangle \over V_{\rm p}} &=&\int \dr p \dr {\bf
k} ~ \f{1}{2 E_p}\f{1}{\cosh(\f{2\pi}{\cQ} E_p)
+\cosh(\f{2\pi p}{\cQ})}\f{2\sinh(\f{2\pi p}{\cQ})}{\sinh(\pi p\cQ)} \nonumber \\
&=& \int \dr p \,\dr{\bf k} \,\dr E \, \f{\theta(E)
\delta(E^2-p^2-{\bf k}^2 -M^2)}{\cosh(\f{2\pi}{\cQ}E)
+\cosh(\f{2\pi p}{\cQ})}\f{2\sinh(\f{2\pi p}{\cQ})}{\sinh(\pi p\cQ)}
\nonumber \ . \eea
Now, decompose the spatial momentum as ${\bf k} =(\ell, {\bf
k}_\bot)$ and consider the boosted frame along the $\ell$-direction
by velocity $\varepsilon$. The phase-space variables of the two
frames are then related by the Lorentz transformations:
\bea E =\gamma (E'-\varepsilon \ell'),\quad
\ell=\gamma(\ell' -\varepsilon E') \ . \nonumber \eea
So, the average number density is now expressible in the boosted
frame as
\bea {\langle N(M) \rangle \over V_{\rm p}} &=& \int \dr p \dr {\bf
k}_\bot \dr \ell' \dr E' ~ \f{\theta(E')~
\delta(E^{\prime 2}-\ell^{\prime 2} -p^2- {\bf k}_\bot^2
-M^2)}{\cosh\Big[\f{2\pi}{\cQ}\gamma (E' -\varepsilon
\ell')\Big]
+\cosh(\f{2\pi p}{\cQ})}\f{2\sinh(\f{2\pi p}{\cQ})}{\sinh(\pi p\cQ)}
\nonumber \\
&=& \int \dr p \dr {\bf k}_\bot \dr \ell' ~ \f{1}{2E_p'}
\f{1}{\cosh\Big[\f{2\pi}{\cQ}\gamma (E'_p -\varepsilon
\ell')\Big] +\cosh(\f{2\pi p}{\cQ})}\f{2\sinh(\f{2\pi
p}{\cQ})}{\sinh(\pi p\cQ)} \ , \nonumber \eea
where $E'_p = \sqrt{p^2+ ({\bf k}_\bot^2+\ell^{\prime
2})+M^2}$. In the T-dual picture, this last expression is
interpretable as the average number of closed string emitted out of
electrified Dp-brane {\it in the $R \rightarrow 0$ limit}.
Explicitly, it is given by
\bea {\langle N(M) \rangle \over V_{\rm p}} = \lim_{R \to 0} \frac{R}{2}
\sum_{w=-\infty}^{+\infty} \int \dr p \dr {\bf k}_\bot ~
\f{1}{2E_p'} \f{1}{\cosh\Big[\f{2\pi}{\cQ}\gamma (E'_p -{1
\over 2} \varepsilon R w)\Big] +\cosh(\f{2\pi
p}{\cQ})}\f{2\sinh(\f{2\pi p}{\cQ})}{\sinh(\pi p\cQ)}\ , \label{final}
\eea
where we T-dualized the longitudinal momentum $\ell'$ to the winding
${1 \over 2} Rw$. Notice that the compactification limit $R \rightarrow
0$ is imperative since, for finite $R$, the Lorentz symmetry is
broken in the original picture and the effect of the electric flux
appears in the T-dualized picture. The expression involves sum over
all winding quantum numbers, $w$. However, for any finite value
of the radiation energy $E_p'$, it is evident from the kinematics that $w$ would
take virtually a continuous value in the limit $R \rightarrow 0$.
Thus, in the compactification limit, one can evaluate the
phase-space sum and integral via saddle-point approximations. It is obvious from
Eq.(\ref{final}) that the saddle-point for $\omega$ is located around a value set by
the electric field $\varepsilon$ times the energy $E_p'$. It is large for a large value
of the transverse mass $M$. Notice also that the integrand depends
nontrivially on the winding quantum number $w$ and also that the integrand at zero
winding $w = 0$ is exponentially suppressed compared to that at the saddle point.

In addition to the average number density, one also
would like to extract spectral shape of the emitted closed strings.
Convolution of the spectral
shape is measured by moments of the energy emitted $\langle E^N(M)
\rangle/V_{\rm p}$ and also by moments of the winding number emitted $\langle \omega^N
(M) \rangle / V_{\rm p}$ ($N=1,2,\cdots$). They are extracted by weighing $E_{p,w}^N$ or
$\omega^N$ into the phase-space integral of Eq.(\ref{final}):
\bea
{\langle E^N(M) \rangle \over V_{\rm p}} &=& \lim_{R \to 0} \frac{R}{2} \!\!
\sum_{w=-\infty}^{+\infty} \int \dr p \dr {\bf k}_\bot ~
\f{1}{2E_p'} (E_p')^N~\f{1}{\cosh\Big[\f{2\pi}{\cQ}\gamma (E'_p -{1
\over 2} \varepsilon R w)\Big] +\cosh(\f{2\pi
p}{\cQ})}\f{2\sinh(\f{2\pi p}{\cQ})}{\sinh(\pi p\cQ)} \nonumber \\
{\langle w^N(M) \rangle \over V_{\rm p}} &=& \lim_{R \to 0}
\frac{R}{2} \!\! \sum_{w=-\infty}^{+\infty} w^N \int \dr p \dr {\bf
k}_\bot ~ \f{1}{2E_p'} \f{1}{\cosh\Big[\f{2\pi}{\cQ}\gamma (E'_p -{1
\over 2} \varepsilon R w)\Big] +\cosh(\f{2\pi
p}{\cQ})}\f{2\sinh(\f{2\pi p}{\cQ})}{\sinh(\pi p\cQ)}. \ \
\label{observable} \eea
These observables transform covariantly under the Lorentz boost, so those
measured in the rest frame can be related to those measured in the boosted frame.

\subsect{Direct computation: winding versus no-winding }

Though the Lorentz covariance argument makes it clear that the
emitted closed strings carry nonzero winding quantum number $w$ set by the electric
field $\varepsilon$, for those who would like to see it explicitly, we shall
evaluate spectral observables Eq.(\ref{observable}) directly using the exact
boundary states constructed in the previous section. In particular, we can show
explicitly that radiation of closed strings with zero winding is
exponentially suppressed due to the presence of the electric field.
This phenomenon is the geometric counterpart of the same phenomenon
for the rolling tachyon
\cite{Mukhopadhyay:2002en,Lambert:2003zr,Karczmarek:2003xm,Nagami:2003yz}
qualitatively, but there is some quantitative difference as we will
see.

The starting point of our argument is the boundary states of the
electrified D-brane
\bea | B, \varepsilon \rangle = \int_{0}^\infty \dr p
\int_{-\infty}^{\infty} \dr \omega ~ \Psi_\varepsilon (p,\omega)
|p,\omega;\varepsilon \rangle\rangle \ , \nonumber \eea
where Ishibashi states $| p,\omega;\varepsilon \rangle\rangle $ can
be constructed by successive operations of T-duality and Lorentz boost as in
section \ref{3}. The boundary wave function
$\Psi_\varepsilon(p,\omega)$ represents the coupling to zero modes
and, for NS sector,
\bea \Psi^{\sNS}_\varepsilon (p,\omega) = \frac{i \sqrt{2}\cQ
\sinh(\frac{ 2 \pi p}{\cQ})}
{2\cosh[\frac{\pi}{\cQ}(p+\gamma\omega)]\cosh[\frac{\pi}{\cQ}(p-\gamma\omega)]}
\cdot \frac{\Gamma(i\cQ p)\Gamma(1+i\frac{2p}{\cQ})}
{\Gamma(\frac{1}{2}-i\frac{\gamma\omega}{\cQ}+i\frac{p}{\cQ})
\Gamma(\frac{1}{2}+i\frac{\gamma\omega}{\cQ}+i\frac{p}{\cQ})} \ . \nonumber
\eea
We focus on the NS sector wave function, since other sectors are readily obtainable by
the spectral flow. As was discussed in \cite{Nakayama:2004yx}, the emission rate does
not show any qualitative difference among these different spin structures (at
least in the UV region).

For a fixed transverse mass $M$, the average number density of emitted closed string is
given by
\bea {\langle N (M) \rangle \over V_{\rm p}} &=& \int \dr {\bf
k}\int_{0}^{\infty} \frac{\dr p}
{2 E_p}~\gamma^{-2}|\Psi^{\rm NS}_\varepsilon(p, E_p)|^2 \nonumber \\
&=& \int \dr {\bf k} \int_{0}^{\infty} \frac{\dr
p}{2E_p} ~ \gamma^{-2}\frac{2\sinh\left(\frac{2\pi p}{\cQ}\right)}
{\left[ \cosh\left(\frac{2\pi \gamma E_p}{\cQ}\right) +
\cosh\left(\frac{2\pi p}{\cQ}\right)\right]\sinh(\pi \cQ p)} \
,\label{emit} \eea
where $E_p = \sqrt{p^2 + {\bf k}^2+ M^2} $ is the on-shell energy of
the emitted closed string states. Likewise, $N$-th moment of the
energy emitted $\langle E^N (M) \rangle/ V_{\rm p}$ is obtainable by
weighing the integrand by $E_p^N$.

These spectral moments involve sum over all final states\footnote{One
might observe that coupling to higher oscillator modes are different
level by level, which could invalidate our naive use of only the
zero mode boundary wave function. However this can be avoided by
taking a suitable gauge. Alternatively, one may evaluate the
imaginary part of the one-loop amplitude in the Minkowski signature, from the emission rate
may be extracted via the optical theorem
\cite{Karczmarek:2003xm,Nagami:2003yz}. Physically, these two prescriptions should
yield identical result.}. It is evident from Eq.(\ref{emit}) that the
phase-space integral involves sum over the infinite massive states and hence may cause
ultraviolet catastrophe. To see if the divergence
actually occurs, we need to sum over these massive modes
for large values of $M$. Evaluating it via the saddle point method, we find that
\begin{eqnarray}
{\langle N (M) \rangle \over V} &\simeq& \int \dr {\bf
k} \int_0^{\infty} \frac{\dr p}{M}\,
\gamma^{-2}e^{\left(\frac{2\pi}{\cQ}-\pi \cQ\right)p -
\frac{2\pi}{\cQ} \gamma\sqrt{p^2+{\bf k}^2+M^2}} \nonumber \\
&\simeq&  e^{-2\pi M\gamma
\sqrt{1-\frac{\cQ^2}{4}+\varepsilon^2(\frac{1}{\cQ}-\frac{\cQ}{2})^2
}}\  . \label{asym}
\end{eqnarray}
Now, density of states of the emitted closed string for a given mass $M$ is
the same as that of the open string. For large $M$, it grows as $ n(M) \sim
M^{-3}e^{2\pi M \sqrt{1-\frac{\cQ^2}{4}}}$ \cite{Kutasov:1990sv}.
So, summing over the transverse mass $M$, the average number density of all closed strings
emitted is estimated as
\begin{eqnarray}
\frac{\langle N \rangle}{V_{\rm p}} &=& \int_0^{\infty} \dr M e^{2\pi M
\sqrt{1-\frac{\cQ^2}{4}}} \int \dr
{\bf k} \int_0^{\infty} \frac{\dr p}{M^4}
~\gamma^{-2}e^{\left(\frac{2\pi}{\cQ}-\pi \cQ\right)p -
\frac{2\pi}{\cQ} \gamma\sqrt{p^2+{\bf k}^2+M^2}} \nonumber \\
&\simeq& \gamma^{-2}\int^{\infty} {\dr L}~
L^{\frac{5-p}{2}-3}e^{-\pi L(\gamma-1)} \nonumber \\
&\simeq& \gamma^{-2}(\gamma-1)^{\frac{p-1}{2}} \ \label{saddle},
\end{eqnarray}
where in the second line we have introduced a dummy `radial'
variable $L^2 = (2/{\cal Q})^2 (M^2 + {\bf k}^2 + p^2)$ and performed
the angular integration first. The last integration is ultraviolet finite.
We also note that all the spectral moments are infrared finite since the boundary wave
function in the NS-sector does not have a pole in the $p, E \to 0$ limit.

It would be illuminating to compare the result Eq.(\ref{saddle})
with that for the decay of an unstable D-brane via tachyon rolling in the flat
spacetime. In the latter situation, the transition
probability is given by $|\Psi_\varepsilon(\omega)|^2 \sim
\frac{1}{\sinh^2\pi \gamma\omega}$. It results in the average number density as
\bea \frac{\langle N \rangle}{V} \simeq \int^{\infty} \dr M ~
e^{-2\pi M(\gamma-1) } \ . \nonumber \eea
Notice that this is manifestly ultraviolet finite but may have infrared
divergence. Apparently, effects of electrifying the D-brane is
qualitatively the same but quantitatively different --- response to
$\varepsilon$ is somewhat different in the spectral weight in
Eq.(\ref{asym}). This is attributed to nontrivial saddle point of
the integrand in Eq.(\ref{asym}). In contrast,
the average total number density of closed strings Eq.(\ref{saddle}) exhibits similar
behavior as that for rolling tachyon case. However, notice that
since the contribution from the higher level is exponentially suppressed in saddle-point
approximation, the change of summation into integration is not trustful in the
evaluation of the average total number density.\footnote{This is to be contrasted against
the consideration of \cite{Chen:2004vw}.}

Several comments are in order.
\begin{itemize}
    \item For bare D-brane decay, spectral density
    of the closed strings emitted was independent of the number of NS5-branes
    $N = {2 /\cQ^2}$. This ensured that, regardless of the number of NS5-branes, the
    spectral moments of emitted closed string are exactly same as those of tachyon
    rolling in flat spacetime. However, such `geometric equivalence' is lost
    at quantitative level once the D-brane is electrified.

    \item Although it is outside the scope of our interest, we have exactly the same
    behavior as for the tachyon rolling if we formally set $\cQ = \sqrt{2}$. As pointed
    out earlier, this case corresponds to the decay of the rolling D-brane in a conifold
    background, not in the NS5-brane background. If we formally extend $\cQ$ below the
    selfdual point, Eqs.(\ref{asym}, \ref{saddle}) is absolutely ultraviolet convergent even for the bare
    D-brane case, $\varepsilon = 0$. Given that the Hagedorn density of state makes sense
    only for $\cQ < 2$, the relevant range for the convergence would be $\sqrt{2} \le
    \cQ \le 2$.

        \item Our result implies that negligible number of the closed string and negligible
        portion of the Dp-brane energy is emitted to zero-winding states. Of course, this is
        as expected from the Lorentz covariance argument in the previous subsection (See also
        \cite{toappear}). In the case of the rolling tachyon, \cite{Sen:2003xs} also argued
        that, after compactification, all the energy of electrified Dp-brane is actually
        radiated away into closed strings with nonzero winding quantum number. See also
        \cite{Gutperle:2004be} for an explicit check to this effect.
\end{itemize}

The last point is what we already explained intuitively in the previous subsection
by appealing to the Lorentz invariance, so we shall now supplement it by direct
computations. Compactify along the direction of the electric field to a circle of radius $R$.
Then the average number density for a fixed mass $M$ is now modified to
\bea {\langle N(M) \rangle \over V_{\rm p}} \simeq \sum_{w=-\infty}^{+\infty} \int \dr
{\bf k} \int_0^{\infty} \frac{\dr p}{M}\,
\gamma^{-2}e^{\left(\frac{2\pi}{\cQ}-\pi \cQ\right)p -
\frac{2\pi}{\cQ} \gamma\left(\sqrt{p^2+{\bf k}^2+M^2+({1 \over 2}
w R)^2}- {1 \over 2} \varepsilon Rw\right)} \ \label{Mvalue}
\eea
by summing over the winding mode contributions ($w = 0$
sector reduces to Eq.(\ref{asym})). For large value of $M$, this sum over the winding
modes can be evaluated by the saddle-point method. The result is
\bea {\langle N(M) \rangle \over V_{\rm p}} \simeq \int \dr {\bf k} \int_0^{\infty}
\frac{\dr p}{M^{\frac{1}{2}}}\,e^{\left(\frac{2\pi}{\cQ}-\pi
\cQ\right)p - \frac{2\pi}{\cQ}\sqrt{p^2+{\bf k}^2+M^2}} \ . \label{fina} \eea
Notice that dependence on $\varepsilon$ in the exponent has disappeared after saddle-point
sum over the winding quantum number $w$.
Integrations over $p$, ${\bf k}$ and finally over $M$ can be done exactly the same way as in
\cite{Nakayama:2004yx}. This leads to power-like ultraviolet catastrophe for the
average number density:
\bea \frac{\langle N \rangle}{V_{\rm p}} \simeq \int^{\infty} \dr M ~
M^{-\frac{p-1}{2}-1}\ , \nonumber \eea
or for higher spectral moments of the energy or the winding number. Notice further
that the dependence on ${\cal Q}$ in Eq.(\ref{fina}) has disappeared after integration
over $p, {\bf k}$.

The main point we would like to bring out is that proper account of the
winding modes recovers universal feature shared by both the tachyon rolling and
the radion rolloing of D-branes. This result asserts that the closed string emitted
mainly consists of the highly winding ones. As discussed in the previous subsection, this
result is also what one would expect from elementary consideration of Lorentz covariance
of spectral observables. The result also partially answers the puzzle mentioned in the first
comment: after the inclusion of the winding modes, quantitative
difference between the rolling tachyon and the rolling D-brane disappeared simply because
spectral observables no longer depend on $\cQ$ or $\varepsilon$. We emphasize again that
dependence on $\cQ$ and $\varepsilon$ disappeared as a consequence of sum over the winding
quantum number.

\section{Conclusion and Discussion}\label{6}
In this paper, we studied rolling radion dynamics of electrified
D-brane in the NS5-brane background. We confirmed that ``geometric
realization" of tachyon rolling in terms of radion rolling continues
to hold to situations where the D-brane is electrified. We
constructed exact boundary states of rolling radion and compared the
results with the effective field theory analysis. In the
supergravity limit $N = {2/\cQ^2} \to \infty$, we confirmed that the
two approaches fully agree with each other.

For bare D-brane, it was shown that the decay of rolling radion is the
same as that of the rolling tachyon {\it irrespective of} the total number of
NS5-brane $\cQ = \sqrt{\frac{2}{N}}$. In this context, it is worthwhile pointing out
that the effective action agrees with each other for a specific
number of the NS5-brane --- $\cQ = \frac{1}{\sqrt{2}}$ for the
bosonic rolling tachyon and $\cQ =1$ for the supersymmetric rolling
tachyon.

 After electrifying the D-brane, we found that exponential
 suppression of the spectral density of the emitted closed string,
 which is qualitatively analogous to the rolling tachyon dynamics of
 electrified D-brane, actually begins to {\it depend} on the
 value of $\cQ$, or the number of the NS5-branes. This can be understood
 as a quantitative difference between the rolling D-brane and the rolling
 tachyon dynamics. It would be interesting to give a holographic dual
 explanation of this fact (in the context of little string theory) or to study
 distinguishing role of the deformed conifold background where the
 dependence on $\varepsilon$ exactly coincides with the rolling tachyon
 problem. However, we also found that this quantitative difference
 may be an artifact of the noncompact space: after proper inclusion of the
 winding modes, the universal property of the decaying process is recovered.
 This completely agrees with the situation in the rolling tachyon.

To conclude, we would like to address a natural
question on the fate of the radion rolling for electrified D-brane.
Our analysis shows that conserved currents consist of two parts --- the
pressureless ``tachyon (radion) matter" and the fundamental string
fluid, just as in the case of the rolling tachyon. From the
geometrical viewpoint, on the other hand, we expect that the rolling
D-brane will make a bound state with the NS5-brane. Thus it would be
interesting to understand what happens to the fundamental string charge, in
addition to the Ramond-Ramond charge, originally carried by the D-brane. To answer
this nonperturbative question, neither the effective field theory analysis nor the
boundary states analysis seems to be adequate. At the least, it calls for quantum
counterpart of the boundary states. On the other hand, the open string completeness conjecture
(see \cite{Sen:2004nf} and references therein) might indicate that our analysis has
already given a clue to this issue. The dual little string theory description or
the M-theoretic consideration should also play a relevant role here.

\vskip.2in \noindent
\section*{Acknowledgment}
We thank Changrim Ahn, Massimo Bianchi, Yasuaki Hikida, Oleg Lunin,
Augusto Sagnotti, Yuji Sugawara and Jung-Tay Yee for useful
discussions. YN is supported in part by a Grant for 21st Century COE
Program ``QUESTS'' from the Ministry of Education, Culture, Sports,
Science, and Technology of Japan. KP was supported in part by
I.N.F.N., by the E.C. RTN programs HPRN-CT-2000-00122 and
HPRN-CT-2000-00148, by the INTAS contract 99-1-590, by the
MURST-COFIN contract 2001-025492 and by the NATO contract
PST.CLG.978785. SJR was supported in part by the National Science
Foundation under Grant No. PHY99-07949, by MOST-KOSEF under Leading
Scientist Grant, and  by the Alexander von Humboldt Foundation
through W.F. Bessel research award. HT is supported in part by JSPS
Research Fellowships for Young Scientists.

\end{document}